\documentclass[12pt,english,aps,manuscript]{article}
\usepackage[T1]{fontenc}
\usepackage[latin1]{inputenc}
\usepackage{geometry}
\usepackage{amsmath}
\usepackage{amssymb}
\usepackage[numbers]{natbib}

\makeatletter

\usepackage{geometry}

\makeatletter

\usepackage{color}

\makeatletter
\newcommand{\bee}{\begin{equation}}
\newcommand{\ee}{\end{equation}}
\newcommand{\beea}{\begin{eqnarray}}
\newcommand{\eea}{\end{eqnarray}}

\makeatother

\makeatother

\makeatother

\usepackage{babel}
\begin{document}

\section*{~~~~~~~~~~~~~~~~~~~~~~~~~~~~~~~~~~~~~~~~~~~~~~~~~~~~~~~~~~~~~~~~~~~~~~~~~~~~~~~~~~~~~~~~~~~~~~~~\textmd{\normalsize COLO-HEP-567}}

\begin{center}
\textbf{\Large A Unified Approach to Supersymmetry Breaking }
\par\end{center}{\Large \par}

\begin{center}
\vspace{0.3cm}
 
\par\end{center}

\begin{center}
{\large S. P. de Alwis$^{\dagger}$ } 
\par\end{center}

\begin{center}
Physics Department, University of Colorado, \\
 Boulder, CO 80309 USA 
\par\end{center}

\begin{center}
\vspace{0.3cm}
 
\par\end{center}

\begin{center}
\textbf{Abstract} 
\par\end{center}

General formulae for the soft SUSY breaking terms, valid in any SUGRA
context, were derived in the mid-nineties. Since SUSY is not expected
to have quantum anomalies, they should be valid in the quantum theory
and be RG invariant down to the soft SUSY breaking scale. This observation
enables us to give a uniform treatment of all phenomenological models
for SUSY breaking and transmission, such as AMSB, GMSB, etc. In particular
we find that the much discussed RG invariant formulae for soft SUSY
breaking parameters in AMSB, effectively depend on a strong assumption
of factorizability of the matter Kaehler metric. We then argue that
there is no necessity for having ad hoc constructions such as mAMSB
to counteract the negative squared slepton mass problem, since the
natural framework that emerges in a sequestered model is one in which
gaugino masses are as in AMSB, and the other soft terms are generated
by RG running as in gaugino mediation. 

\begin{center}
\vspace{0.3cm}
 
\par\end{center}

\vfill{}

$^{\dagger}$ dealwiss@colorado.edu

\eject

\section{Introduction}

A theory of supersymmetry (SUSY) breaking necessarily involves supergravity
(SUGRA). The reason is that spontaneous supersymmetry breaking in
a global SUSY theory leads to a positive vacuum energy at the scale
of supersymmetry breaking. Given the limits on superpartner masses
this is necessarily many orders of magnitude greater than the observed
value of the cosmological constant (CC). In SUGRA by contrast the
CC can in principle be fine-tuned to values parametrically below the
SUSY breaking scale.

However SUGRA is not an ultra-violet complete theory. It must necessarily
be contained within a consistent theory of quantum gravity. It is
widely expected that string theory is such a theory. In any case whether
or not string theory is the UV completion of SUGRA, it is clear that
some more fundamental theory must replace SUGRA at some scale $\Lambda\leq M_{P}$
the Planck scale, where even the notion of a smooth metric background
is expected to breakdown. The theory below such a scale may be expanded
in terms of the number of derivatives scaled by $\Lambda$.

General formulae for the so-called soft terms which characterize the
supersymmetry breaking parameters in the MSSM have been discussed
many years ago \citep{Kaplunovsky:1993rd,Brignole:1997dp}. As was
stressed in the first of these references these formulae are renormalization
group invariant at least down to the soft SUSY breaking scale. While
they were originally discussed within the context of string phenomenology,
they are valid regardless of the validity of string theory. The necessary
criteria are the following:
\begin{itemize}
\item There is some UV scale (say $\Lambda<M_{P}$) up to which there is
a low energy (4D) SUGRA description of nature. 
\item At the relevant energy scale $E,\, E/\Lambda\ll1$, the SUGRA theory
can be truncated to a two derivative theory. For consistency this
also requires the restriction $F/\Lambda^{2}\ll1$. In string theory
based models for example, $\Lambda\lesssim M_{KK}$ where $M_{KK}$
is the lowest Kaluza-Klein mass.
\item Under these conditions the effective theory - including quantum corrections
can be described by a real analytic Kaehler potential $K$, an analytic
superpotential $W$ and an analytic gauge coupling function $f$. 
\item The set of chiral superfields in the theory can be broken up into
two classes. 1) A set of (gauge neutral) fields which will have large
ground state values at the minimum of the scalar potential, and may
have non-zero F-terms, and 2) a set of fields which will have essentially
zero vacuum expectation values (vev's) and F-terms. Here large means
values that are at least a significant fraction of $\Lambda$. Actually
in all but GMSB these values will be at least of order $M_{P}$. Fields
in category 1) will be referred to as moduli (even though they need
not have anything to do with the moduli of string theory).
\end{itemize}
In GMSB type models the SUSY breaking sector has nothing to do with
the string theory moduli. It is usually taken to be some O'Raifeartaigh
type model. Such a model may be put in a canonical form, with heavy
fields which do not participate in SUSY breaking, and a single (light)
field $X$ which develops an F-term \citep{Komargodski:2009jf}. If
the theory is embedded in a string theory it is first necessary to
integrate out all the closed string moduli in a supersymmetric Minkowski
background, so that one can have a SUGRA formulation at some energy
scale lower than the mass of the lightest closed string modulus. (see
for example \citep{deAlwis:2010ud}). GMSB has in addition to this
SUSY breaking sector, a messenger sector (in direct mediation models
it may be part of the breaking sector), that couples to both the SUSY
breaking sector and the standard model gauge fields. The mass scale
of the messengers is (essentially) given by $X_{0}$, the vev of the
SUSY breaking field. Thus the consistency of the two derivative theory
below the messenger scale then requires that $\frac{F^{X}}{X_{0}}|\ll1$%
\footnote{This also means that the component calculations of gaugino and scalar
masses in GMSB should only be trusted at lowest non-trivial order
in this parameter since the starting point has effectively ignored
the higher order terms.%
}.

A bottom up approach to soft mass formulae focussing mostly on the
quantum effects associated with gauge mediated supersymmetry breaking
(GMSB) was published \citep{ArkaniHamed:1998kj} several years after
\citep{Kaplunovsky:1993rd}. Here the supersymmetry breaking was introduced
as a non-dynamical (or spurion) field with a non-zero F-term. One
of the main points of our paper, is to emphasize not only that these
arguments are implicit in \citep{Kaplunovsky:1993rd,Brignole:1997dp},
but also that the formulae given there, give a more powerful method
of deriving all the quantum effects. In particular it shows how all
of the currently popular models of SUSY breaking emerge in a natural
fashion from different assumptions about the moduli dependence of
the matter metric.

In the rest of the paper we will focus on two derivative terms under
the assumption that the supersymmetry breaking is small in the above
sense that $F/\Lambda^{2}\ll1$, where $\Lambda$ is the mass of the
lowest scale that has been integrated out. In the next section we
will discuss the general framework of two derivative SUGRA theories
and then how different popular models of SUSY breaking emerge from
this general framework. In particular we will discuss the validity
of AMSB arguments. Then we will discuss how GMSB fits into this framework.
Finally we will discuss the natural replacement of AMSB within the
context of sequestered theories. This is the mechanism which has been
called gaugino anomaly mediation (inoAMSB), discussed in \citep{deAlwis:2009fn,Baer:2010uy}.
We end with a brief summary of our results.

\section{Supergravity formalism and the soft SUSY breaking terms }

The most general manifestly supersymmetric action for chiral scalar
fields $\Phi$ coupled to supergravity and gauge fields, when restricted
to no more than two derivatives, can be expressed in terms of three
functions. i) The (real analytic) Kaehler potential $K(\Phi,\bar{\Phi})$,
the analytic superpotential $W(\Phi)$, and the (analytic) gauge coupling
function (or functions if there is more than one simple group factor)
$f(\Phi)$. (see for example \citep{Wess:1992cp},\citep{Gates:1983nr}). 

The action of superspace supergravity coupled to chiral scalars and
Yang-Mills fields is (following the notation and conventions of \citep{Wess:1992cp}
but with $M_{P}=\kappa^{-1}=1$),

\begin{eqnarray}
S & = & -3\int d^{6}z2{\cal E}(-\frac{\bar{\nabla}^{2}}{4}+2R)\exp[-\frac{1}{3}K(\Phi,\bar{\Phi};Q,\bar{Q}e^{2V})]+\nonumber \\
 &  & \left(\int d^{6}z2{\cal E}[W(\Phi,Q)+\frac{1}{4}f(\Phi){\cal W}^{a}{\cal W}^{a}]+h.c.\right).\label{eq:actionWB}
\end{eqnarray}
 The arguments given in \citep{Wess:1992cp},\citep{Gates:1983nr}
(or for that matter the original arguments of \citep{Cremmer:1982en})
imply that the effective action at some scale after including all
effects (classical and quantum) coming from integrating out states
at higher scales, must still be of this form. In particular it remains
an action that is determined by three functions $K,W,f$ which are
just dependent on the physical chiral fields $\Phi$. All that can
change (relative to some 'classical' expression) is the functional
form of these superfields. 

The main point of this paper is that any low energy physical effect
should be obtainable from this action as long as the restrictions
of the supersymmetric derivative expansion discussed in the introduction
are satisfied. This means that once the functional form of $K,W$
and $f$ are given (including the quantum corrections) one should
be able to read off the physical masses and couplings of the theory
(at the scale at which we expect these forms to be valid), from the
expression in component form for the above action that is given in
(for instance) Appendix G of \citep{Wess:1992cp}.

\subsection{General Expressions for soft terms and RG invariance}

What is of most interest for us in the context of (low energy) SUSY
breaking is the boundary values of the soft masses and couplings,
which in the context of the MSSM will become the parameters of phenomenological
interest. The theory above has a set of gauge neutral fields $\Phi=\{\Phi^{A}\}$,
which in a string theory context for instance, would be identified
as the moduli determining the size and shape of the internal 6D manifold
as well as the string coupling. In general we need to find the point
at which these are stabilized in a SUSY breaking fashion, and is such
that none of the charged fields $Q=\{Q^{a}\}$ get a vacuum value.
If one finds such a minimum then the soft masses are obtained in the
manner described below \citep{Kaplunovsky:1993rd}.

We expand the superpotential and the Kaehler potential in powers of
the charged fields, i.e. we write
\begin{eqnarray}
W & = & \hat{W}(\Phi)+\frac{1}{2}\tilde{\mu}_{ab}(\Phi)Q^{a}Q^{b}+\frac{1}{6}\tilde{Y}_{abc}(\Phi)Q^{a}Q^{b}Q^{c}+\ldots,\label{eq:Wexpn}\\
K & = & \hat{K}(\Phi,\bar{\Phi})+Z_{a\bar{b}}(\Phi,\bar{\Phi})Q^{a}Q^{\bar{b}}+[X_{ab}(\Phi,\bar{\Phi})Q^{a}Q^{b}+h.c.]+\ldots.\label{eq:Kexpn}
\end{eqnarray}
Then one may easily compute the soft masses from the well known expression
for the scalar potential in supergravity and get \citep{Kaplunovsky:1993rd}\citep{Brignole:1997dp},
\begin{eqnarray}
(m^{2})_{a}^{a'} & \equiv & Z^{a'\bar{b}}m_{a\bar{b}}^{2}=\frac{1}{3}(2V_{0}+F^{A}F^{\bar{B}}\hat{K}_{A\bar{B}})\delta_{a}^{a'}-F^{A}F^{\bar{B}}R_{A\bar{B}a\bar{b}}Z^{a'\bar{b}}\label{eq:softmass}\\
 & = & \frac{1}{3}(2V_{0}+F^{A}F^{\bar{B}}\hat{K}_{A\bar{B}})\delta_{a}^{a'}-F^{A}F^{\bar{B}}\partial_{A}(Z^{a'\bar{b}}\partial_{\bar{B}}Z_{a\bar{b}})\label{eq:softmass1}
\end{eqnarray}
Note that these expressions are written for the canonically normalized
fields so that these expressions are valid for the normalized squared
mass matrix. Note also that while the first term is proportional to
the unit matrix, the second is not necessarily so, and hence is in
general a potential source of flavor violation.

Now while \eqref{eq:actionWB} is manifestly (off-shell) supersymmetric,
the component form (given for instance in Appendix G of \citep{Wess:1992cp})
only has on-shell supersymmetry. In fact in arriving at the latter
a series of (super) Weyl transformations and field redefinitions of
chiral multiplets has been performed. This is necessary in order to
get to the Einstein frame for (super) gravity and Kaehler normalization
for the chiral fields (with for instance the scalar field kinetic
term being of the form $K_{a\bar{b}}\partial_{\mu}\phi^{b}\partial^{\mu}\bar{\phi}^{\bar{b}}$).
Now in the quantum theory these transformations do not leave the measure
invariant and there is an anomaly. However as usual this anomaly just
changes the gauge coupling function (at the two derivative level),
and has no effect on the Kaehler metric. Hence the above formula \eqref{eq:softmass}
remains valid in the quantum theory - assuming of course that the
appropriate $K,W,f$ are used. For instance the dilaton component
of the Kaehler potential of the heterotic string is a term of the
form $K\sim-\ln(S+\bar{S})$. Due to string loop effects this term
gets changed to $K\sim-\ln(S+\bar{S}-\Delta(M,\bar{M})/16\pi^{2})$
\citep{Derendinger:1991hq}. This will obviously change the curvature
term in \eqref{eq:softmass} - but this has nothing to do with an
anomaly. Similar considerations apply to the expressions for the $\mu$,
$B\mu$ and $A$ terms given in \citep{Kaplunovsky:1993rd}\citep{Brignole:1997dp}. 

The gauge coupling function on the other hand does experience an anomaly,
since the above field redefinitions give contributions to the measure
which are of the form $\exp\{\#\int\tau{\cal W}{\cal W}+h.c.)$, where
$\tau$ is some (chiral) superfield transformation parameter. In particular
this means that the gauge coupling function $g_{phys}(\Phi,\bar{\Phi})$and
the gaugino mass $M(\Phi,\bar{\Phi})$ (after supersymmetry breaking
i.e. $F^{A}\ne0$ for at least one value of $A$), are given by 
\begin{eqnarray}
\frac{1}{g_{phys}^{2}} & = & \Re f+\frac{c}{16\pi^{2}}\hat{K}|_{0}-\sum_{r}\frac{T(r)}{8\pi^{2}}\ln\det{\bf Z}^{(r)}|_{0}+\frac{T(G)}{8\pi^{2}}\ln\frac{1}{g_{phys}^{2}},\label{eq:gboundary}\\
\frac{2M}{g_{{\rm phys}}^{2}} & = & (F^{A}\partial_{A}f+\frac{c}{8\pi^{2}}F^{A}\hat{K}_{A}-\sum_{r}\frac{T_{a}(r)}{8\pi^{2}}F^{A}\partial_{A}\ln\det{\bf Z}^{(r)}|_{0})\times(1-\frac{T(G_{a})}{8\pi^{2}}g_{{\rm {\rm phys}}}^{(a)2})^{-1}.\label{eq:mboundary}
\end{eqnarray}

It should be stressed that the three formulae \eqref{eq:softmass}\eqref{eq:gboundary}\eqref{eq:mboundary}
for the soft mass (and analogous formulae for the $\mu,B\mu$ and
$A$ terms), the gauge coupling and gaugino mass, are all expressions
valid at whatever scale the explicit expressions for the Kaehler potential
$K$ as a function of the moduli $\Phi$ is given. Thus if $K$ is
obtained from string theory (after incorporating $\alpha'$ and string
loop corrections), one expects these expressions to be valid at some
point close to the string scale. These formulae are then to be used
as the boundary conditions for renormalization group (RG) evolution.
To one loop order the RG evolution values of the coupling function
and the gaugino mass at some scale $\mu$, would be given in terms
of the value at the ($\Phi$ independent) boundary scale $\Lambda$
($\lesssim M_{string}$ if the fundamental theory is string theory),
by making the replacements 
\begin{equation}
f\rightarrow f+(b/16\pi^{2})\ln(\Lambda^{2}/\mu^{2})\label{eq:fRG}
\end{equation}
 and $ $$g_{phys}^{-2}\rightarrow f$ on the RHS of \eqref{eq:gboundary}.
Note that to this order the second equation is unchanged and in fact
the factor in parenthesis in the last term on the RHS can be replaced
by unity. However the above formulae can actually be interpreted as
being valid to all orders in the loop expansion, provided in addition
to the replacement for $f$ \eqref{eq:fRG} one replaces $g_{phys}^{2}\rightarrow g_{phys}^{2}(\mu^{2})$
and 
\begin{equation}
{\bf Z}^{(r)}(\Phi,\bar{\Phi})\rightarrow{\bf Z}^{(r)}(\Phi,\bar{\Phi};g^{2}(\mu)).\label{eq:ZRG}
\end{equation}
Thus we have the following formulae for the parameters at the infra
red RG scale $\mu$, which are expected to be valid to all orders
in the loop expansion:

\begin{eqnarray}
\frac{1}{g_{phys}^{2}}(\Phi,\bar{\Phi};\mu) & = & \Re f+\frac{b}{16\pi^{2}}\ln\frac{\Lambda^{2}}{\mu^{2}}+\frac{c}{16\pi^{2}}\hat{K}|_{0}-\sum_{r}\frac{T(r)}{8\pi^{2}}{\rm tr}\ln{\bf Z}^{(r)}(g^{2}(\mu))|_{0}\nonumber \\
 &  & +\frac{T(G)}{8\pi^{2}}\ln\frac{1}{g_{phys}^{2}(\mu)},\label{eq:gmu}\\
\frac{2M}{g_{{\rm phys}}^{2}}(\Phi,\bar{\Phi};\mu) & = & (F^{A}\partial_{A}f+\frac{c}{16\pi^{2}}F^{A}\hat{K}_{A}-\sum_{r}\frac{T_{a}(r)}{8\pi^{2}}F^{A}\partial_{A}{\rm tr}\ln{\bf Z}^{(r)}(g^{2}(\mu))|_{0})\nonumber \\
 &  & \times(1-\frac{T(G_{a})}{8\pi^{2}}g_{{\rm {\rm phys}}}^{(a)2}(\mu))^{-1}.\label{eq:mmu}
\end{eqnarray}
The first of these equations is the integrated form of the NSVZ beta
function \citep{Novikov:1985rd} with the boundary condition (at $\mu=\Lambda$)
fixed by the KL supergravity correction (the third term above)%
\footnote{Note that the original derivation in \citep{Kaplunovsky:1994fg} for
the last term in \eqref{eq:gmu}, used the NSVZ beta function. However
as shown in \citep{ArkaniHamed:1997mj} (see also \citep{deAlwis:2008aq}),
this term can be derived by arguments very similar to those used in
\citep{Kaplunovsky:1994fg} to get the third and fourth terms.%
}. In fact differentiating with respect to $t\equiv\ln\mu$ and assuming
the gauge neutrality of the moduli and gravitational interactions
so that $\Phi,\hat{K}$ are independent of $t$ %
\footnote{This assumption is violated in GMSB like theories where the SUSY breaking
sector is coupled via the messenger sector to the standard model gauge
group.%
} we get the NSVZ equation for the exact beta function: 
\begin{equation}
\frac{\beta}{g}=\frac{g^{2}}{16\pi^{2}}\frac{b+2\sum_{r}T(r){\rm tr}\gamma_{r}}{(1-g^{2}(\mu)T(G)/8\pi^{2})},\label{eq:NSVZ}
\end{equation}
where we've defined 
\begin{equation}
\gamma_{r}=\frac{1}{2}\frac{d}{dt}\ln{\bf Z}^{(r)}(\Phi,\bar{\Phi};g^{2}(\mu)).\label{eq:gamma}
\end{equation}
\eqref{eq:mmu} does not seem to have been written down explicitly
before - though it is of course a straightforward consequence of the
KL formula. Note its resemblance to the NSVZ beta function. Also differentiating
\eqref{eq:mmu} with respect to $t$ gives the beta function for the
gaugino mass
\begin{equation}
\beta_{M}\equiv\frac{dM}{dt}=\frac{2M\beta-g^{3}\sum_{r}T(r){\rm tr}\gamma^{(1)}/8\pi^{2}}{g(1-g^{2}T(G)/8\pi^{2})},\label{betaM}
\end{equation}
where
\begin{equation}
\gamma_{r}^{(1)}=F^{A}\partial_{A}\gamma_{r}\label{eq:gamma1}
\end{equation}
This equation has been derived earlier in \citep{Hisano:1997ua}\citep{Jack:1997pa},
but under additional assumptions. Note that these two equations are
compatible with the (expected) relation 
\begin{equation}
\frac{dM}{dt}=-F^{A}\partial_{A}\frac{\beta}{g},\label{eq:dMdt}
\end{equation}
which follows from the RG invariance of the moduli i.e. 
\begin{equation}
[\frac{d}{dt},F^{A}\partial_{A}]=0,\label{eq:RGinvariance}
\end{equation}
 and the formula $F^{A}\partial_{A}(1/g^{2})=2M/g^{2}$ (see for example
\citep{Wess:1992cp}). This may be rewritten as, 
\begin{equation}
F^{A}\partial_{A}g=-Mg.\label{eq:Fdg}
\end{equation}

What these supergravity considerations tell us, is to take the values
(at the high scale $\Lambda)$ for the scalar masses given by \eqref{eq:softmass},
along with the analogous formulae for the $\mu,B\mu$ and $A$ terms
as well as the formulae \eqref{eq:gboundary}\eqref{eq:mboundary}
for the gauge couplings and gaugino mass, as boundary values for integrating
the RG equations of the MSSM, to find the values at the scale $\mu$.
This follows the standard practice used for mSUGRA for instance. This
is consistent with the formula \eqref{eq:softmass} once we modify
the Kaehler metric for the matter fields appropriately.

Thus consider the one loop correction to the matter metric. Keeping
the functional form in terms of the moduli fields in the standard
one loop counter term calculation in the supersymmetric theory (see
for example \citep{Drees:2004jm}), we have (putting $g^{2}=1/\Re f$)
\begin{equation}
\Delta Z_{a\bar{b}}=\frac{1}{16\pi^{2}}\ln\frac{\Lambda^{2}}{\mu^{2}}(Y_{acd}\bar{Y}_{\bar{b}\bar{c'}\bar{x'}}Z^{c\bar{c}}Z^{d\bar{d'}}-2(\Re f)^{-1}T(r)Z_{a\bar{b}}).\label{eq:1loopcounter}
\end{equation}
When this correction is made to the metric in formula \eqref{eq:softmass},
one gets additional one loop terms when the derivatives with respect
to the moduli act on moduli dependent functions in \eqref{eq:1loopcounter}.
In particular the derivatives acting on the gauge coupling function
will give contributions proportional to the gaugino mass squared:
\begin{equation}
-F^{A}F^{\bar{B}}\Delta R_{A\bar{B}a\bar{b}}\sim-F^{A}F^{B}\partial_{A}\partial_{B}\Delta Z_{a\bar{b}}\sim\frac{1}{16\pi^{2}}\ln\frac{\Lambda^{2}}{\mu^{2}}T(r)g^{6}|F^{A}\partial_{A}f|^{2}+\ldots=\frac{1}{16\pi^{2}}\ln\frac{\Lambda^{2}}{\mu^{2}}T(r)g^{2}|M|^{2},\label{eq:gauginoterm}
\end{equation}
where $M$ is the gaugino mass (to lowest order). This gives the term
$\beta^{(m^{2})}\sim g^{2}|M|^{2}T(r)/16\pi^{2}$ in the beta function
for the (squared) scalar mass which drives the gaugino mediated contribution
to it.

\subsection{Gravity (moduli) mediation }

The generic and most natural situation that arises in supersymmetry
breaking (given that we need to start from SUGRA), is what is often
called gravity mediated SUSY breaking. However since the actual transmission
of supersymmetry breaking is through the gauge neutral fields that
we have called moduli, it is more appropriate to call this moduli
mediated SUSY breaking (MMSSB). In general this will lead to flavor
changing neutral currents (FCNC) at a level which is disallowed by
experiment, so some additional assumptions are needed. In mSUGRA (see
for example \citep{Arnowitt:2009qt} for a recent review) one makes
the assumption of universal scalar masses and a universal A-term.
This follows from a certain factorization assumption for the metric
on MSSM field space and the assumption of independence of the Yukawa
couplings from the SUSY breaking moduli. If the cut-off (beyond which
the SUGRA needs to be replaced by string theory) is well below the
Planck scale, then these assumptions are preserved by quantum corrections%
\footnote{For a recent discussion of these issues see \citep{deAlwis:2010sw}.%
}.

\subsubsection{Sequestered moduli mediation: AMSB}

Here we will focus on the so-called sequestered models which give
a different approach to solving the FCNC problem within MMSB. This
has been called anomaly mediated supersymmetry breaking (AMSB) \citep{Randall:1998uk,Giudice:1998bp}.
The only anomaly that one has in a SUGRA consistently coupled to an
anomaly free gauge theory (like the standard model), is the Weyl anomaly
discussed in \citep{Kaplunovsky:1994fg}. For the gaugino mass this
gives an extra term (namely the second term on the RHS in \eqref{eq:mboundary}
or\eqref{eq:mmu}).%
\footnote{The question of whether there is yet another {}``anomaly'' term
proportional to the gravitino mass as is claimed in much of the phenomenological
literature has been addressed and answered in the negative in \citep{deAlwis:2008aq}.
The essence of the argument there was that any such claim implies
that quantum effects break supersymmetry. Subsequent to the publication
of that paper several authors have claimed (in effect) that even if
it is absent in the Wilsonian action it will be present in the 1PI
action. These arguments will be addressed in a separate publication.%
} 

As for the scalar masses and the $A$ and $B\mu$ terms, expressions
for them are derived in the AMSB literature by assuming that quantum
corrections that are proportional to $\ln\Lambda/\mu$ should be replaced
by $\ln\Lambda C/\mu$ where $C$ is the Weyl compensator whose F-terms
are then given a non-zero value. As shown in \citep{deAlwis:2008aq}
if indeed $C$ is the Weyl compensator this procedure will actually
violate the Weyl invariant formalism, and in fact the compensator
will become a propagating field. On the other hand one may regard
this insertion as a spurion, in which case what we have is an explicit
breaking of SUSY. In any case one gets an elegant set of formulae
which have the added benefit of being RG invariant. The relations
in question are (see for example \citep{Jack:1999aj} where these
are shown to be RG invariant); 
\begin{eqnarray}
M & = & M_{0}\frac{\beta}{g},\label{eq:MRGI}\\
(m^{2})_{\, b}^{a} & = & -|M_{0}|^{2}\frac{d\gamma_{\, b}^{a}}{dt}=-|M_{0}|^{2}\beta\frac{d\gamma_{\, b}^{a}}{dg},\label{eq:scalarRGI}\\
A_{abc} & = & -M_{0}\frac{dY_{abc}}{dt},\label{eq:ARGI}\\
B_{ab} & = & -M_{0}\frac{d\mu_{ab}}{dt}.\label{eq:BRGI}
\end{eqnarray}
Here $M_{0}$ is a constant mass parameter. For completeness, we have
included the expressions for the $A$ and $B$ terms, but let us just
focus on the first two equations. Firstly we see that identifying
$M_{_{0}}=F^{\phi}$ we have the AMSB formula for the scalar mass.
In the original version of AMSB this is identified with $m_{3/2}$,
the gravitino mass.

Let us now show that these formulae have nothing to do with the Weyl
anomaly. Writing
\begin{equation}
F^{A}\partial_{A}=F^{A}\partial_{A}|_{g}+F^{A}\partial_{A}g\partial_{g}|_{\Phi},\label{eq:FpartialA}
\end{equation}
 we see that 
\[
F^{A}\partial_{A}\beta=\frac{2\sum_{r}T(r)F^{A}\partial_{A}|_{g}{\rm tr}\gamma}{1-g^{2}T(G)/8\pi^{2}}+F^{A}\partial_{A}g\partial_{g}|_{\Phi}\beta,
\]
where we've used \eqref{eq:NSVZ} in the first term. Now if this term
can be ignored we see that (using \eqref{eq:Fdg} $F^{A}\partial_{A}\beta=-Mgd\beta/dg$),
\begin{equation}
\frac{d}{dt}\left(\frac{Mg}{\beta}\right)=0\Rightarrow\left(\frac{Mg}{\beta}\right)=M_{0},\label{eq:AMSBRG1}
\end{equation}
where $M_{0}$ is a RG invariant constant. This is precisely the AMSB
formula (now derived without inserting Weyl compensator fields) but
with the above assumption. For the scalar mass we get (for simplicity
we work with one family of matter fields $Q$), after tuning the CC
to zero,
\begin{eqnarray}
m^{2} & = & m_{3/2}^{2}-F^{\bar{A}}\partial_{A}|_{g}F^{B}\partial_{B}|_{g}\ln Z(\Phi,\bar{\Phi};g)\nonumber \\
 &  & -(\bar{M}g\partial_{g}|_{\Phi}F^{B}\partial_{B}|_{g}+F^{\bar{A}}\partial_{A}|_{g}Mg\partial_{g}|_{\Phi})\ln Z(\Phi,\bar{\Phi};g)\nonumber \\
 &  & -2\bar{M}g\partial_{g}|_{\Phi}\frac{Mg}{\beta}\gamma,\label{eq:m2all}
\end{eqnarray}
where we've put $\gamma=\frac{1}{2}d\ln Z/dt$ and used $\partial_{g}|_{\phi}=\beta^{-1}d/dt$.
Using \eqref{eq:AMSBRG1} we then have for the last line,
\begin{equation}
\Delta m^{2}=-2\frac{|M|^{2}g^{2}}{\beta}\frac{d\gamma}{dg}|_{\Phi}=-2\beta|M_{0}|^{2}\frac{d\gamma}{dg}|_{\Phi}.\label{eq:AMSBscalar}
\end{equation}
This is precisely the AMSB formula for the scalar masses! However
clearly this is not a necessary consequence of sequestering (otherwise
sequestering would be disastrous). The first line for example may
be set equal to zero in sequestered models, but this can be done only
at some fixed UV scale (say $\Lambda$) where the last factor in the
second term has its classical value. Below this scale this will change
and then this line will give non-zero contributions. The second line
will also in general be non-zero even in sequestered models. Finally
we had to use \eqref{eq:AMSBRG1} to arrive at \eqref{eq:AMSBscalar}.

The key assumption that would lead to both \eqref{eq:AMSBRG1} and
\eqref{eq:AMSBscalar} is the factorization of the kinetic function
for matter fields i.e. $Z(\Phi,\bar{\Phi};g(\mu))=Z_{0}(\Phi,\bar{\Phi})Z_{1}(g(\mu))$
\textit{and} (for the scalar mass case) that the classical contribution
is negligible (sequestering). This factorization assumption will lead
to the second line of \eqref{eq:m2all} becoming zero, and the second
term on the first line becoming independent of $g$. So if the classical
contribution (i.e. the value at the UV scale) is zero (i.e. sequestering)
then the first line is also zero, thus giving the so-called AMSB expression
\eqref{eq:AMSBscalar} as the sole contribution. 

Similar arguments can be made for the $A$ and $B\mu$ terms.

\subsubsection{Sequestered moduli mediation: inoAMSB}

As we argued above sequestering alone does not lead to the AMSB formulae.
Additional assumptions are needed and in fact they lead to disastrous
consequences as is well known. What then is the most natural moduli
mediated scenario, once the low energy theory inherited from the ultra
violet theory (string theory) is of the sequestered form? Our claim
is that this is the one discussed in \citep{deAlwis:2009fn,Baer:2010uy}.
It may be viewed as the correct form of what is traditionally known
as AMSB. Thus in this case the gaugino mass is essentially given by
the anomaly terms in the KL formula, since the classical contribution
is highly suppressed (sequestered). The classical scalar masses (i.e.
the mass at the UV scale $\Lambda$), are all negligible due to sequestering.
The masses at the low energy (gravitino) scale are then generated
by RG running and acquire non-zero values (with positive definite
values for squarks and sleptons) while giving the necessary electro-weak
breaking contributions to the Higgs mass matrix by RG running. The
dominant contribution to this running is what has been called the
gaugino mediation mechanism \citep{Kaplan:1999ac,Chacko:1999mi}.

The string theory basis for inoAMSB, was discussed in \citep{deAlwis:2009fn}
and its phenomenological consequences in \citep{Baer:2010uy}, so
we will not pursue it further here. The main point that we wish to
emphasize is that regardless of its origins in string theory, given
a sequestered scenario, the resulting model of SUSY breaking with
no additional assumptions is inoAMSB.

\subsection{GMSB}

In GMSB typically there is a SUSY breaking sector and a messenger
sector (which in direct mediation would be part of the former), in
addition to the MSSM. If the theory is embedded in string theory,
then in addition there could be a low energy string theoretic moduli
sector as well. However in order to construct a viable GMSB model
(avoiding additional fine tuning), the latter sector should be integrated
out in a (Minkowski) supersymmetric fashion (\citep{deAlwis:2010ud}).
Furthermore the lowest modulus mass should be greater than the scale
of the SUSY breaking sector. Although the corresponding string vacua
appear to be rather sparse, in principle this is achievable. In such
a situation we can focus on a SUSY breaking sector characterized by
some UV scale $\Lambda\ll M_{P}$, which would be the mass of the
lightest modulus that has been integrated out. The potential in the
SUSY breaking sector is then expected to have a (meta-stable) minimum
with $X=X_{0}\ll\Lambda$ and $|F^{X}|=\sqrt{3}m_{3/2}M_{P}\ne0$,
where the first relation comes from tuning the CC to zero after SUSY
breaking. The SUSY breaking field $X$ couples to the messengers through
a superpotential term of the form $\Delta W=\lambda Xf\tilde{f}$.
In principle there may be additional SUSY breaking fields

In such a situation the KL formula \eqref{eq:gmu} for the gauge coupling
function undergoes a modification below the messenger scale due to
the messenger threshold (\citep{Giudice:1997ni}:

\begin{eqnarray}
\frac{1}{g_{phys}^{2}}(\Phi,\bar{\Phi};\mu) & = & \Re f+(\frac{b_{>}}{16\pi^{2}})\ln(\frac{\Lambda^{2}}{X\bar{X}})+(\frac{b_{<}}{16\pi^{2}})\ln(\frac{X\bar{X}}{\mu^{2}})+\frac{c}{16\pi^{2}}\hat{K}|_{0}-\sum_{r}\frac{T(r)}{8\pi^{2}}{\rm tr}\ln{\bf Z}^{(r)}(g^{2}(\mu))|_{0}\nonumber \\
 &  & +\frac{T(G)}{8\pi^{2}}\ln\frac{1}{g_{phys}^{2}(\mu)},\label{eq:gmu-1}\\
\frac{2M}{g_{{\rm phys}}^{2}}(\Phi,\bar{\Phi};\mu) & = & (F^{A}\partial_{A}f+\frac{T_{mess}}{16\pi^{2}}\frac{F^{X}}{X}+\frac{c}{16\pi^{2}}F^{A}\hat{K}_{A}-\sum_{r}\frac{T(r)}{8\pi^{2}}F^{A}\partial_{A}{\rm tr}\ln{\bf Z}^{(r)}(g^{2}(\mu))|_{0})\nonumber \\
 &  & \times(1-\frac{T(G)}{8\pi^{2}}g_{{\rm {\rm phys}}}^{2}(\mu))^{-1}.\label{eq:mmu-1}
\end{eqnarray}
Now the GMSB effect becomes relevant only in situations that the gravitino
mass has been tuned to be several orders of magnitude below the soft/weak
scale. Now all terms on the RHS in the expression for the gaugino
mass \eqref{eq:mmu} except the second are $O(m_{3/2})$, or (in the
case of the $\ln Z$ term on the first line) of higher order in perturbation
theory, while the second term is $O(m_{3/2}M_{P}/X_{0}\mbox{)\ensuremath{\gg}}m_{3/2}$
(since $X_{0}\ll\Lambda$). Thus we have the usual GMSB expression
\begin{equation}
\frac{2M}{g_{{\rm phys}}^{2}}(\Phi,\bar{\Phi};\mu)\simeq+\frac{T_{mess}}{16\pi^{2}}\frac{F^{X}}{X}.\label{eq:MGMSB}
\end{equation}
Note that \eqref{eq:mmu-1} remains valid at all loop order even in
the presence of the messenger threshold. In fact in addition to the
second term on the RHS (which is necessarily a one-loop effect), there
is an additional messenger threshold effect in the third term coming
from the term in the sum when $\Phi^{A}=X$. In this case (if $m_{3/2}$
is tuned to be sufficiently small) the messenger threshold gives the
dominant contribution to this sum;
\begin{equation}
\sum_{r}\frac{T(r)}{8\pi^{2}}F^{A}\partial_{A}{\rm tr}\ln{\bf Z}^{(r)}(g^{2}(\mu))|_{0}\simeq\frac{T_{mess}}{8\pi^{2}}F^{X}\partial_{X}{\rm tr}\ln{\bf Z}^{(r)}(g^{2}(\mu))=2\frac{T_{mess}}{8\pi^{2}}\frac{F^{X}}{X}(\gamma(\mu)-\gamma(X))\frac{\beta_{>}-\beta_{<}}{\beta_{<}}|_{X}.\label{eq:higherorder}
\end{equation}
Note that this term is of higher loop order than \eqref{eq:MGMSB}.
For $ $$\gamma=c\alpha/4\pi$ it is a factor $c\alpha\ln(X/\mu)/16\pi^{3}$
smaller than the leading contribution \eqref{eq:MGMSB}. In may be
of relevance in situations where the leading contribution vanishes
\citep{Komargodski:2009jf} as in some direct mediation models.

Similar arguments can be applied to the soft masses. Thus we may write,
again using\eqref{eq:FpartialA} in \eqref{eq:softmass1} 
\begin{eqnarray}
m^{2} & = & m_{3/2}^{2}-F^{\bar{A}}\partial_{A}|_{g}F^{B}\partial_{B}|_{g}\ln Z(\Phi,\bar{\Phi};g)\nonumber \\
 &  & -(F^{\bar{A}}\frac{\partial g}{\partial\Phi^{\bar{A}}}\partial_{g}|_{\Phi}F^{B}\partial_{B}|_{g}+F^{\bar{A}}\partial_{A}|_{g}F^{B}\frac{\partial g}{\partial\Phi^{B}}\partial_{g}|_{\Phi})\ln Z(\Phi,\bar{\Phi};g)\nonumber \\
 &  & -F^{\bar{A}}\frac{\partial g}{\partial\Phi^{\bar{A}}}\partial_{g}|_{\Phi}F^{B}\frac{\partial g}{\partial\Phi^{B}}\partial_{g}|_{\Phi}\ln Z(\Phi,\bar{\Phi};g).\label{eq:m2all-1}
\end{eqnarray}
Note that now because of the existence of the messenger threshold,
we cannot any longer just use \eqref{eq:AMSBRG1}. When the sum over
$A$ gets to $\Phi^{A}=X$ (i.e. is one of the fields which break
SUSY and couples to the messengers) one gets additional contributions
from the last term in \eqref{eq:m2all-1}. For $\mu\rightarrow X_{0}$
this contribution is given by the standard GMSB expression, 
\begin{equation}
m^{2}\sim2|\frac{F^{X}}{X}|^{2}(\beta_{>}-\beta_{<})\frac{\partial\gamma}{\partial g}\simeq2|\frac{F^{X}}{X}|^{2}T_{mess}c\left(\frac{\alpha}{4\pi}\right)^{2}.\label{eq:m2GMSB}
\end{equation}
 In the last relation we've used the one loop beta function and anomalous
dimension $\gamma=c\alpha/4\pi$. This term is of course always positive
and is the standard GMSB formula. The point that we wish to emphasize
here is that unlike in the case of AMSB, here the suppression of gravity
mediated effects (effectively $F^{X}/X_{0}\sim m_{3/2}M_{P}/X_{0}\mbox{\ensuremath{\gg}}m_{3/2}$)
means that all other terms in \eqref{eq:m2all-1} are suppressed compared
to \eqref{eq:m2GMSB}. Thus with this one assumption (namely $m_{3/2}\ll M_{W}$)
we get the GMSB formulae from the general framework.

\section{Conclusions}

We have given a unified treatment of currently popular phenomenological
models of supersymmetry breaking, within the framework of the general
theory developed in the nineties. The main point of our analysis is
that these general arguments are valid even at the quantum level,
provided we take into account the necessary corrections to the supergravity
potentials ($K,W)$ and the gauge coupling function $f$. This enables
us to understand where the so-called AMSB formulae come from. These
arguments also show that the natural consequence of sequestering,
is the mediation mechanism that has been called inoAMSB. Finally we
showed how the GMSB formulae emerge from this framework, and pointed
out an additional term that has not been generally discussed in the
literature.

\section{Acknowledgments}

The research of SdA is partially supported by the United States Department
of Energy under grant DE-FG02-91-ER-40672.

\section*{}

\[
\]
 \bibliographystyle{apsrev} \bibliographystyle{apsrev}
\bibliography{myrefs}

\end{document}